\documentclass{article}

\usepackage{amsmath, amsthm, amssymb}
\usepackage{graphicx}
\usepackage{verbatim}
\usepackage{natbib}
\usepackage{caption}
\usepackage{subcaption}
\usepackage{fancyvrb}
\usepackage{enumerate}
\usepackage{relsize}

\usepackage{hyperref}
\usepackage[margin=1.5in]{geometry}
\hypersetup{colorlinks,citecolor=blue,urlcolor=blue,linkcolor=blue}

\usepackage{stefan_tex}
\graphicspath{{./figures/}}

\newcommand\blfootnote[1]{%
  \begingroup
  \renewcommand\thefootnote{}\footnote{#1}%
  \addtocounter{footnote}{-1}%
  \endgroup
}

\newcommand{\hCV}{\widehat{\operatorname{CV}}}

\theoremstyle{plain}
\newtheorem{prop}{Proposition}

\theoremstyle{definition}

\newtheorem{assu}{Assumption}

\theoremstyle{remark}

\author{Stefan Wager \\ Stanford University}
\date{}
\title{Cross-Validation, \\ Risk Estimation, and Model Selection}

\begin{document}

\maketitle

\begin{abstract}
Cross-validation is a popular non-parametric method for evaluating the accuracy of
a predictive rule. The usefulness of cross-validation depends on the task we want to employ
it for. In this note, I discuss a simple non-parametric setting, and find that cross-validation is
asymptotically uninformative about the expected test error of any given predictive rule, but allows for
asymptotically consistent model selection. The reason for this phenomenon is that the leading-order
error term of cross-validation doesn't depend on the model being evaluated, and so cancels out when
we compare two models. This note was prepared as a comment on a paper by Rosset and Tibshirani,
forthcoming in the Journal of the American Statistical Association.
\end{abstract}

How best to estimate the accuracy of a predictive rule has been a longstanding question\blfootnote{\hspace{-6.5mm}
I am grateful for several helpful conversations with Brad Efron. This work was supported by
National Science Foundation grant DMS-1916163 and a Facebook Faculty Award.}
in statistics. Approaches to this task range from simple methods like Mallow's Cp to algorithmic
techniques like cross-validation; see \citet{arlot2010survey}, \citet{efron1983estimating,efron2004estimation},
\citet*{hastie2009elements}, \citet{mallows1973some}, and references therein.
\citet{rosset2018fixed} contribute to this discussion by considering how some classical results on the ``optimism''
of the apparent error of a predictive rule, i.e., the amount by which the training set error of a fitted statistical predictor
is expected to underestimate its test set error, change when we consider a random- versus fixed-$X$ sampling design.
This is a welcome addition to the literature as, in modern statistical settings, we often need to work with large observational
datasets that were incidentally collected as a by-product of some other task, and in these cases random-$X$ modeling
is more appropriate than the classical fixed-$X$ approach.

There are two reasons a statistician may want to estimate the accuracy of a predictive model. One is
to simply understand the quality of its predictions: For example, a company may need to choose whether to purchase a new
forecasting tool, and want to evaluate its accuracy in order to better understand the value of the tool for its
business. Another motivation is model selection: Cross-validation and related methods are often used to choose
between competing predictive rules, or to set the complexity parameter with methods like the lasso
\citep*{chetverikov2016cross,hastie2009elements}. For the first task, we in fact need to accurately \emph{estimate}
the accuracy of the predictive rule itself, and the results of \citet{rosset2018fixed} are focused on this task. For the
second, however, we only need to \emph{compare} the accuracy of two competing rules; and this statistical task
ends up having fairly different properties than risk estimation.

In this note, I compare properties of $K$-fold cross-validation for both risk estimation and model comparison under random-$X$
asymptotics. We have access to independent and identically distributed samples $(X_i, \, Y_i) \in \xx \times \RR$, and want to
predict $Y_i$ from $X_i$ under squared error loss. The optimal predictive rule is the conditional response function $\mu(x) = \EE{Y \cond X = x}$.
For simplicity, I'll focus on evaluating models \smash{$\hmu(x)$} whose root-mean-squared excess error \smash{$\mathbb{E}[(\hmu(X) - \mu(X))^2]^{1/2}$}
decays with sample size $n$ as $n^{-\gamma}$, for some exponent $1/4 < \gamma < 1/2$. In other words, I assume that the predictor
converges slower than the parametric $1/\sqrt{n}$ rate, but faster than the $1/\sqrt[4]{n}$.

In this setting, cross-validation
is asymptotically \emph{uninformative} about the test set error of the fitted rule in that, as described more formally below, the analyst would
prefer to estimate the test-set error of \smash{$\hmu(\cdot)$} as \smash{$\operatorname{Err}^* = \mathbb{E}[(Y - \mu(X))^2]$} (which does not depend on the specific
choice of \smash{$\hmu(\cdot)$}) than to use cross-validation. Conversely, cross-validation is asymptotically \emph{consistent} for model
selection, i.e., given the choice of two predictors, it repeatedly picks the more accurate of the two.

Thus, whether we should adopt an optimistic or pessimistic view of cross-validation depends largely on the statistical task at hand.
I want to emphasize that the result presented here is not new; for example, it is implicit in the proof of Theorem 1 of \citet{yang2007consistency}.
The purpose of this discussion is simply to highlight this fact, and to offer a simple argument that applies in the random-$X$ setting
of \citet{rosset2018fixed}.

\paragraph{Numerical example}

\begin{figure}
\begin{center}
\begin{tabular}{cc}
\includegraphics[width=0.45\textwidth]{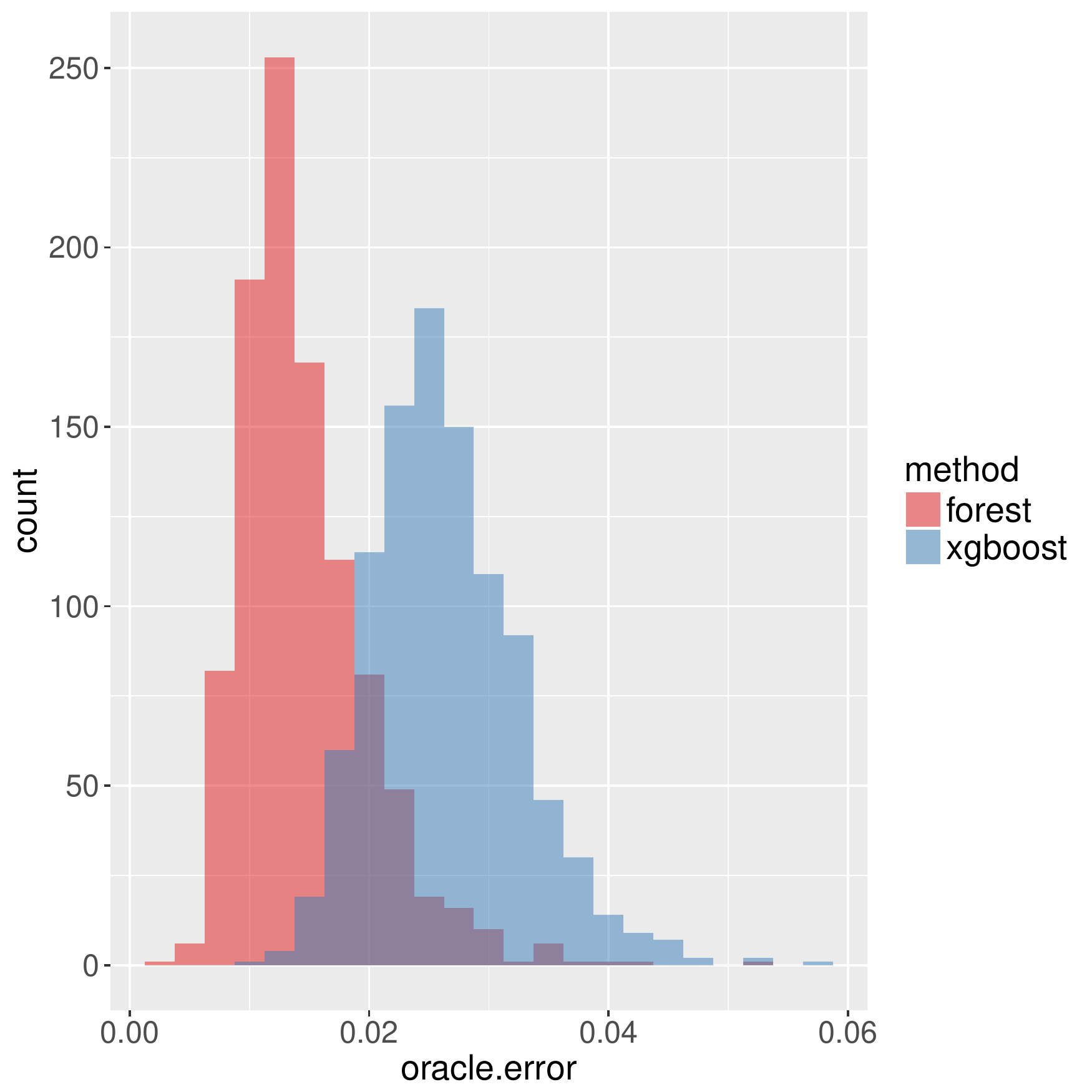} &
\includegraphics[width=0.45\textwidth]{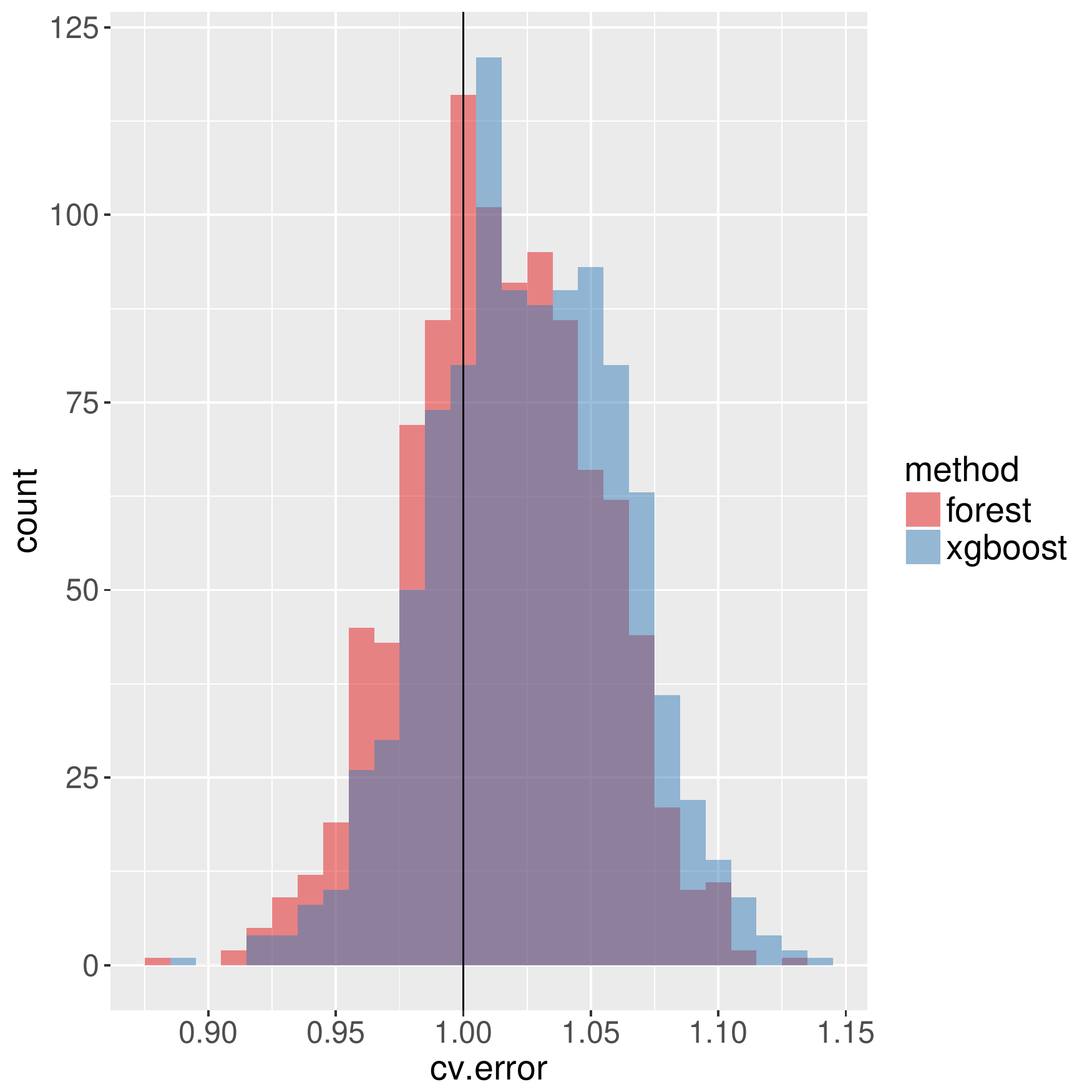} \\
$\text{avg}_\text{test}\sqb{\p{\mu(X_i) - \hmu(X_i)}^2}$ &
$\text{avg}_\text{train}\sqb{\p{Y_i - \hmu^{(-i)}(X_i)}^2}$
\end{tabular}
\caption{Comparison of the performance of boosting and regression forests on the data-generating process
\eqref{eq:dgp}, both in terms of test set excess mean-squared error and training set cross-validation error.
The numbers are aggregated across 1000 simulation replications.}
\label{fig:err}
\end{center}
\end{figure}

Before a more formal discussion, consider the following simulation example. We generate data as
\begin{equation}
\label{eq:dgp}
X \sim \nn\p{0, \, \ii_{p \times p}}, \ \ Y \cond X \sim \nn\p{\mu(X), \, 1}, \ \ \mu(x) = \frac{1\p{\cb{x_1 > 0}}}{1 + e^{-2x_2}},
\end{equation}
with $p = 10$ and $n = 1,600$.
We then seek to fit this signal using gradient boosting as implemented in \texttt{xgboost} \citep{chen2016xgboost}
and regression forests as implemented in \texttt{grf} \citep*{athey2019generalized},
both with built-in parameter tuning.\footnote{For \texttt{grf}, I used the function \texttt{regression\_forest} with
option \texttt{tune.parameters = TRUE}.
For \texttt{xgboost}, I used the function \texttt{cv.xgb}, which cross-validates the number of trees used for boosting. I
set \texttt{nrounds = 1000}, \texttt{early\_stopping\_rounds = 10} and \texttt{max\_depth = 3}, with other parameters set to default.
More extensive cross validation with random search over other parameters, such as \texttt{eta}, \texttt{max\_depth} and \texttt{gamma},
did not improve the performance of boosting here.}
We compare the methods via cross-validation.
For boosting, we use 10-fold cross-validation, whereas for forests we use leave-one-out
(or out-of-bag) evaluation \citep{breiman2001random}.

As shown in the left panel of Figure \ref{fig:err}, forests are noticeably more accurate for this task than
boosting. However, as seen in the right panel, the marginal distribution of the cross-validated errors of
forests versus boosting are nearly indistinguishable. Thus, at first glance, Figure \ref{fig:err} appears to
paint a fairly bleak picture: Forests are more accurate than boosting here, but cross-validation cannot
tell the difference. 

\begin{figure}
\begin{center}
\begin{tabular}{cc}
\includegraphics[width=0.45\textwidth]{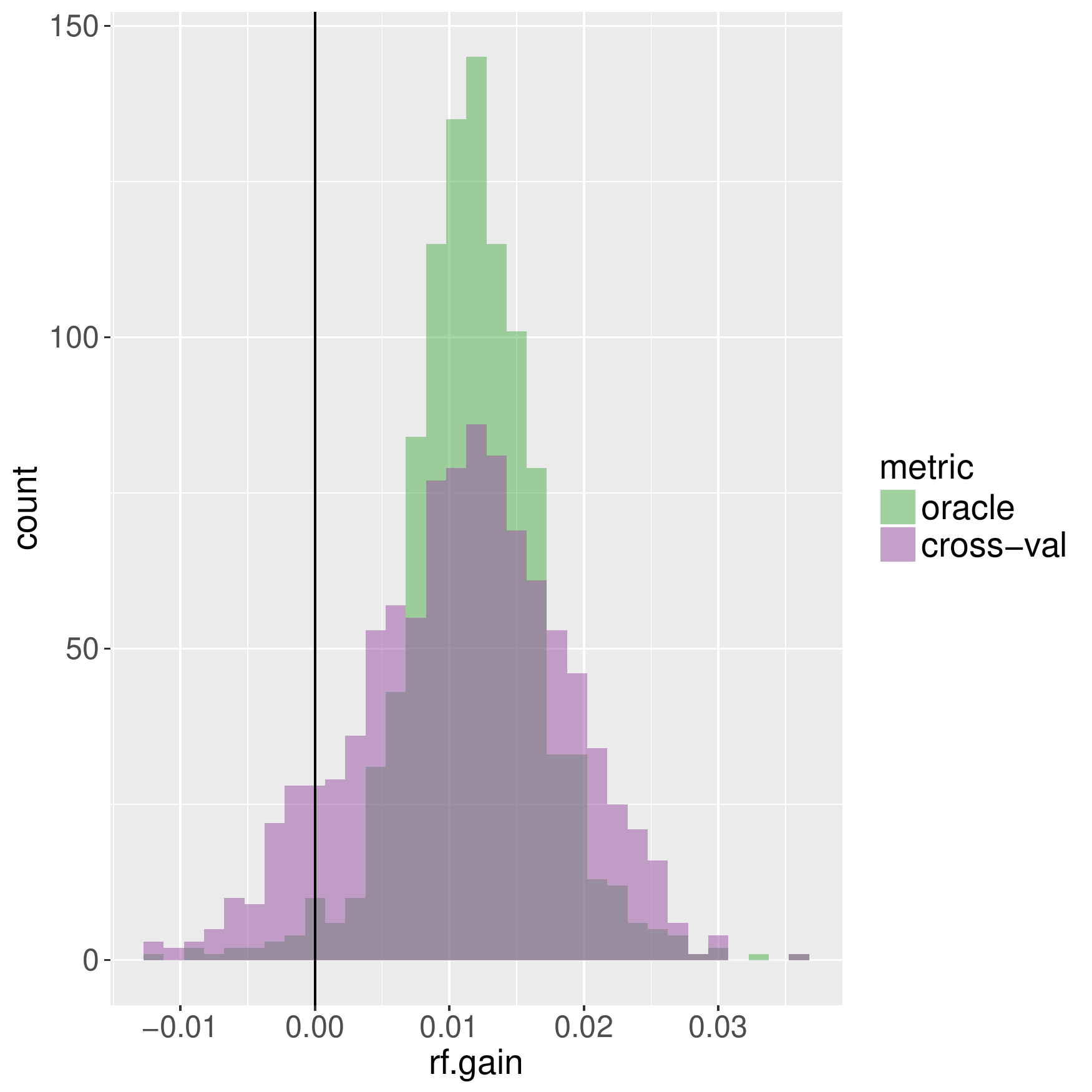} &
\includegraphics[width=0.45\textwidth]{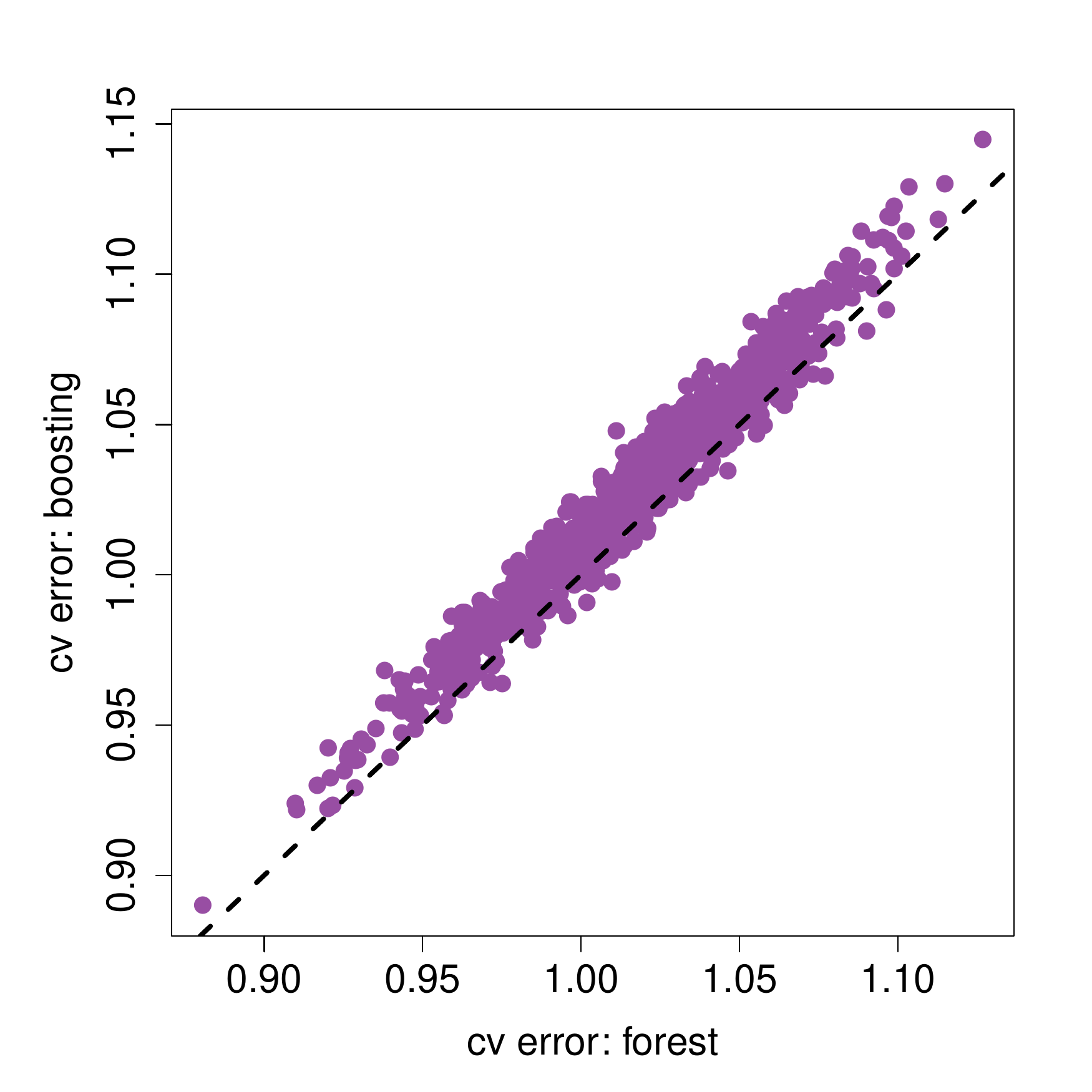} 
\end{tabular}
\caption{In the setting of Figure \ref{fig:err}, comparison of the cross-validation errors of
boosting and forests within a single simulation replication. The left panel shows a histogram of
the cross-validation error of the forest minus the cross-validation error of boosting across the
1000 simulation replications. The overlaid ``oracle'' histogram shows the difference in test set
errors (i.e., differences of the quantities whose marginal histograms are shown in the left panel
of Figure \ref{fig:err}). The right panel shows the same information in a scatter plot.}
\label{fig:cmp}
\end{center}
\end{figure}

As shown in Figure \ref{fig:cmp}, however, the picture clears up when we take differences. As shown
in the left panel of Figure \ref{fig:cmp}, cross-validation consistently picks the more accurate method here.
Accurate model selection is possible despite the finding from Figure \ref{fig:err} because, as shown in
the right panel of Figure \ref{fig:cmp}, the cross-validated error estimates for forests and boosting are
highly correlated, and this shared noise component cancels out when we take a difference.

\paragraph{Formal results}

Cross-validation is used to evaluate an algorithm $A$ that takes a set of $n$ training samples and
turns them into an rule $\hmu(\cdot)$ that predicts $Y_i$ from $X_i$, i.e.,
\begin{equation}
A : \cup_{n = 1}^\infty \cb{\xx \times \RR}^n \rightarrow \cb{\xx \rightarrow \RR}, \ \ \ \
A\p{ \cb{(X_i, \, Y_i)}_{i = 1}^n} = \hmu(\cdot). 
\end{equation}
Some papers use common notation to denote both the algorithm $A$ and the
fitted predictor \smash{$\hmu(\cdot)$}; here, however, it's helpful to disambiguate notation so that we can be specific
about our assumptions. Throughout, we consider cross-validation as a tool for evaluating the algorithm
$A$, rather than the actual predictive rule $\hmu(\cdot)$ that was obtained using the training data.

Given this notation, $K$-fold cross-validation operates as follows. Start with a set of $n$ training samples, and 
divide them into $K$ evenly sized and non-overlapping folds $\set_1, \, ..., \, \set_K$.
Then, for each $k = 1, \, ..., \, K$, run the algorithm $A$ on data in all but the $k$-th fold to obtain $\hmu^{(-k)}(\cdot)$.
The estimate of the error of $A$ when trained on $n$ samples is then
\begin{equation}
\hCV_{n,K}\p{A} = \frac{1}{n} \sum_{k = 1}^K \sum_{i \in \set_k} \p{Y_i - \hmu^{(-k)}(X_i)}^2.
\end{equation}
Throughout, we assume that the method $A$ yields a predictor \smash{$\hmu(\cdot)$} whose
root-mean squared excess test error scales as $n^{-\gamma}$ conditionally on the training data.
This type of behavior can be verified, e.g., for kernel smoothing or local linear regression when $\mu(\cdot)$
satisfies reasonable regularity conditions.

\begin{assu}
\label{assu}
We have access to a stream independent and identically distributed samples $(X_i, \, Y_i) \in \xx \times \RR$ with $\EE{Y^4} \leq \infty$
and $\Var{Y_i \cond X_i} \leq \Omega$.
We also have an algorithm $A$ for learning predictors \smash{$\hmu(\cdot)$} with the following property.
There are constants $0 < C_- \leq C_+ < \infty$ and $0.25 < \gamma < 0.5$
such that, when trained on $n$ samples $\cb{(X_i, \, Y_i)}_{i = 1}^n$, the excess risk of \smash{$\hmu(\cdot)$} scales as
\begin{equation}
\begin{split}
&\limn \PP{n^\gamma \, \EE{\p{\hmu(X) - \mu(X)}^2 \cond \cb{(X_i, \, Y_i)}_{i = 1}^n}^{\frac{1}{2}} \leq C_-} = 0, \\
&\limn \PP{n^\gamma \, \EE{\p{\hmu(X) - \mu(X)}^2 \cond \cb{(X_i, \, Y_i)}_{i = 1}^n}^{\frac{1}{2}} \leq C_+} = 1,
\end{split}
\end{equation}
where $X$ denotes a test sample drawn independently from the training distribution.
\end{assu}

To study the behavior of cross-validation under this assumption, it is helpful to expand-out the square,
as is done in \citet{rosset2018fixed}, \citet{yang2007consistency}, etc.:
\begin{align*}
&\hCV_{n,K}\p{A} = \hCV^*_{n,K} + 2 Z_{n,K} \p{A} + \Delta^2_{n,K}\p{A}, \ \where \\
&\hCV^*_{n,K} = \frac{1}{n}  \sum_{i = 1}^n \p{Y_i - \mu(X_i)}^2, \ \ \ \
\Delta^2_{n,K}\p{A} = \frac{1}{n} \sum_{k = 1}^K \sum_{i \in \set_k}  \p{\mu(X_i) - \hmu^{(-k)}(X_i)}^2, \\
&Z_{n,K} \p{A} = \frac{1}{n} \sum_{k = 1}^K \sum_{i \in \set_k}  \p{Y_i - \mu(X_i)} \p{\mu(X_i) - \hmu^{(-k)}(X_i)}.
\end{align*}
In other words, \smash{$\hCV^*_{n,K}$} is the training set error of the optimal predictor $\hmu(\cdot)$,
\smash{$\Delta^2_{n,K}(A)$} is an oracle estimate of the excess error of the fitted rule, and
\smash{$Z_{n,K}(A)$} is a cross-term.

Given this decomposition, we note that
\begin{equation}
\sqrt{n} \p{\hCV^*_{n,K} - \operatorname{Err}^*} \Rightarrow \nn\p{0, \, \Var{\p{Y- \mu(X)}^2}}, \ \  \operatorname{Err}^* = \EE{\p{Y- \mu(X)}^2}.
\end{equation}
Meanwhile, by our cross-fold construction, we can verify that $\EE{Y_i - \mu(X_i) \cond X_i, \, \hmu^{(-k)}(X_i)} = 0$ for
all $i \in \set_k$, and furthermore that $Y_i - \mu(X_i)$ and $Y_j - \mu(X_j)$ are pairwise uncorrelated for $i, \, j \in \set_k$ conditionally
on \smash{$\hmu^{(-k)}(X_i)$}. The upshot is that, by Assumption \ref{assu}, for all $k = 1, \, ..., \, K$
\begin{align*}
&\EE{\sum_{i \in \set_k}  \p{Y_i - \mu(X_i)} \p{\mu(X_i) - \hmu^{(-k)}(X_i)} \cond \hmu^{(-k)}(\cdot)} = 0 \eqand \\
&\PP{\frac{n_k^{2\gamma}}{\abs{\set_K}}\EE{\p{\sum_{i \in \set_k}  \p{Y_i - \mu(X_i)} \p{\mu(X_i) - \hmu^{(-k)}(X_i)}}^2 \cond \hmu^{(-k)}(\cdot)} \leq C^2_+ \, \Omega} = 1,
\end{align*}
where $n_k = n - \abs{\set_k}$ denotes the amount of training data available to learn \smash{$\hmu^{(-k)}$}. Thus,
by Markov's inequality,
\begin{equation}
Z_{n,K} \p{A} = \oo_p\p{\frac{1}{n^{0.5 + \gamma}}}.
\end{equation}
Finally, given the scaling in Assumption \ref{assu}, the oracle mean-squared excess risk \smash{$\Delta^2_{n,K}(A)$}
scales as $n^{-2\gamma}$ in probability.

A first immediate consequence of this decomposition is that, to first order, the cross-validated error estimate of
$A$ depends only on the test-set error of the optimal predictor $\mu(\cdot)$, and \smash{$\hCV_{n,K}\p{A}$} is asymptotically
equivalent to \smash{$\hCV^*_{n,K}$}. Thus, an analyst wanting to estimate the expected test set error of $\hmu(\cdot)$,
i.e., \smash{$\mathbb{E}[(Y - \hmu(X))^2 \cond \cb{(X_i, \, Y_i)}_{i = 1}^n]$}, under mean squared error would prefer to use a
point estimate \smash{$\operatorname{Err}^* = \mathbb{E}[(Y- \mu(X))^2]$} (which does not depend on \smash{$\hmu$}) than
to use cross-validation.

\begin{prop}
\label{prop:risk}
Under Assumption \ref{assu}, the first-order behavior of \smash{$\hCV_{n,K}\p{A}$} does not depend on the
method $A$ being evaluated:
\begin{equation}
\sqrt{n}\p{\hCV_{n,K}\p{A} - \operatorname{Err}^*} \Rightarrow \nn\p{0, \, \Var{\p{Y- \mu(X)}^2}}.
\end{equation}
\end{prop}

The picture becomes more encouraging, however, when we seek to use cross validation to compare two different predictive
rules. The dominant source of noise \smash{$\hCV^*_{n,K}$} underlying the result in Proposition \ref{prop:risk} does not
depend on $A$, and so cancels out when we compare two rules. Meanwhile, the cross-term $Z_{n,K} \p{A}$ decays faster
than the oracle excess error term \smash{$\Delta^2_{n,K}\p{A}$}, meaning that cross-validation allows for asymptotically
perfect model selection.

\begin{prop}
\label{prop:cmp}
\sloppy{Suppose we have two methods $A$ and $A'$ satisfying \ref{assu} with constants
$(\gamma, \, C_-, \, C_+)$ and $(\gamma', \, C_-', \, C_+')$ respectively. Suppose moreover
that $\gamma > \gamma'$, or that $\gamma = \gamma'$ and $C_+ < C_-'$. Then,}
\begin{equation}
\frac{\hCV_{n,K}\p{A'} - \hCV_{n,K}\p{A}}{\Delta^2_{n,K}\p{A'} - \Delta^2_{n,K}\p{A}} \rightarrow_p 1,
\end{equation}
and $\limn \PP{\hCV_{n,K}\p{A'} > \hCV_{n,K}\p{A}} = 1$.
\end{prop}

Together, Propositions \ref{prop:risk} and \ref{prop:cmp} mean that, given two methods for generating predictive
rules that satisfy Assumption \ref{assu}, the prima facie risk estimates provided by cross-validation are asymptotically independent
of the methods being evaluated, but model selection via cross-validation can accurately pick the better of the
two methods.
Returning to our numerical example presented above, one could argue that
Proposition \ref{prop:risk} predicts the indistinguishability of the two histograms
as observed in the right panel of Figure \ref{fig:err}, whereas Proposition \ref{prop:cmp} helps explain the success of
model selection witnessed in Figure \ref{fig:cmp}.

Estimating the error rate of a prediction rule is an important statistical task, and \citet{rosset2018fixed}
contribute valuable new results to this endeavor in the case of random-$X$ asymptotics. In studying
the properties of cross-validation, though, results are qualitatively different when we focus on model
evaluation versus model comparison. This is not only a formal curiosity, but also affects how we interpret
cross-validation in practical examples; see, e.g., the discussion in Section 2.1 of \citet{nie2017quasi}.
Related facts are also reflected in statistical practice through, e.g., the recommendation to use
McNemar's test to compare the accuracy of two classification rules, or in using a consensus test-train split
for evaluating methods in shared engineering tasks.
It would be interesting to see whether the results of \citet{rosset2018fixed} on optimism allow for
natural extension to the case of model comparison.

\bibliographystyle{plainnat}
\bibliography{references}

\end{document}